\documentclass{ws-procs975x65}

\newcommand{\beq} {\begin{equation}}
\newcommand{\eeq} {\end{equation}}

\begin{document}

\title{Holographic geometries for condensed matter applications}

\author{V.~Ker\"{a}nen and L.~Thorlacius}

\address{Nordita, KTH Royal Institute of Technology and Stockholm University\\
Roslagstullsbacken 23, SE-106 91 Stockholm, Sweden\\
and \\
University of Iceland, Science Institute\\
Dunhaga 3, IS-107 Reykjavik, Iceland\\
E-mail: vkeranen@nordita.org  \  larus@nordita.org}

 \begin{abstract}
Holographic modeling of strongly correlated many-body systems motivates the
study of novel spacetime geometries where the scaling behavior of quantum
critical systems is encoded into spacetime symmetries. Einstein-Dilaton-Maxwell 
theory has planar black brane solutions that exhibit Lifshitz scaling and in some 
cases hyperscaling violation. Entanglement entropy and Wilson loops in the dual 
field theory are studied by inserting simple geometric probes involving minimal 
surfaces into the black brane geometry. Coupling to background matter fields 
leads to interesting low-energy behavior in holographic models, such as U(1) 
symmetry breaking and emergent Lifshitz scaling.
\end{abstract}

\keywords{Black Holes, Gauge Theory - Gravity Correspondence, AdS/CMT.}

\bodymatter

\section{Introduction}

In recent years gauge theory - gravity duality, also referred to as holographic
duality, has emerged as a versatile tool to
investigate strongly coupled dynamics in a variety of physical systems, ranging from the
hydrodynamics of quark-gluon plasma formed in heavy ion collisions to quantum critical
phenomena in condensed matter physics and ultra-cold atomic systems (for reviews 
see {\it e.g.}\cite{DeWolfe:2013cua,Sachdev:2011wg,Adams:2012th}).
The motivation comes from the AdS/CFT conjecture of duality between superstring theory
and maximally supersymmetric Yang-Mills gauge 
theory \cite{Maldacena:1997re,Witten:1998qj,Gubser:1998bc}. The original conjecture and
various refinements of it have passed many nontrivial tests and their validity is by now
accepted by most high-energy theorists, even if it remains unproven. Building on this,
considerable effort has been put into extending the duality to settings where there is no
supersymmetry or even conformal symmetry to constrain the dynamics. This is clearly
more speculative, but may bring us closer to the study of real world systems where
these symmetries are known to be absent. The gravity dual then provides a
phenomenological description of strongly coupled physics on the field theory side, and,
as such, it can be useful even if the underlying dynamics is poorly understood.
The success of this approach will in the end be judged by its ability to match and ultimately 
predict experimental results for strongly coupled systems.

The duality relates gravity in spacetime with a negative cosmological constant to field
theoretic systems, which are {\it a priori\/} far removed from physics in curved spacetime.
More conventional applications of gravitational theory in astrophysics and cosmology
involve vanishing or positive cosmological constant and a spacetime geometry that is
asymptotically flat or de Sitter. Interest in strongly correlated many-body systems thus
motivates the study of novel gravitational solutions that would otherwise be of limited
interest. This includes domain-wall like analogs of black holes with a planar horizon
and carrying various types of matter field `hair'.
In the following we review several holographic constructions that have been developed
to model interesting strong coupling physics. As befits a Marcel Grossman meeting, the
focus will be on the geometries arising in these constructions and various geometric
probes used to study their properties. The presentation will be brief and many important 
topics left out but we hope to give the reader an impression of this new and rapidly 
developing area of application of general relativity. 

A lot of the work on
condensed matter applications of gauge theory - gravity duality has been aimed at
quantum critical systems and we begin our discussion there. In what is usually referred
to as a {\it bottom-up\/} approach, we consider simple gravity models where the scaling
behavior of quantum critical systems is encoded into spacetime symmetries and
many-body interactions are modeled by coupling appropriate matter fields to gravity.
The important issue of to what extent these models may be obtained in a {\it top-down\/}
fashion as low-energy limits of a consistent background in string theory or supergravity
will not be addressed here.

\section{Quantum critical points}
In a quantum phase transition the ground state of a system at zero temperature
changes as a physical parameter (pressure, external magnetic field, etc.) is
varied\cite{Sachdev,Sondhi}. The transition between superconducting and insulating
behavior in certain thin metallic films as a function of their thickness\cite{Haviland}
is a classic example. Another example is provided by the transition to anti-ferromagnetic
order as a function of doping in certain heavy fermion alloys\cite{Lohneysen} and a
quantum critical point may also play a role in explaining the behavior of
high $T_c$ superconductors at low doping.\cite{Sachdev2}.

When a second order quantum critical point is approached, characteristic length scales of
the system diverge and there is an emergent scaling symmetry under
\beq
t\rightarrow \lambda^z\,t , \qquad \vec{x}\rightarrow \lambda \vec{x}.
\label{eq:anisotropic}
\eeq
In a relativistic system energy and distance are inversely related. In this case
the {\it dynamical scaling exponent} is $z=1$ and the scaling symmetry is enhanced to a
conformal symmetry. In non-relativistic systems, on the other hand, the scaling at a quantum
critical point can be asymmetric between the temporal and spatial directions so that $z\neq 1$,
commonly referred to as Lifshitz scaling in the literature.

Finite temperature introduces a length scale that breaks the scaling symmetry at a quantum
critical point. In an otherwise scaling symmetric theory, the temperature dependence of a
characteristic length scale is $\ell\sim T^{-1/z}$. If the corresponding zero temperature system 
is not exactly at the quantum critical point, the temperature dependence exhibits a more 
general scaling form,
\beq
\ell\sim T^{-1/z} \eta(T^{-1/z}\lambda_i),\label{eq:deformed}
\eeq
where $\lambda_i$ are a set of deformation parameters, defined so that they have dimensions
of inverse length and $\lambda_i=0$ corresponds to infinite $\ell$ at $T=0$, and
$\eta$ is some generic function of its arguments.

The systematic study of non-relativistic holographic systems that realize scale symmetry without
conformal symmetry has progressed considerably in recent years and is now at a point where
such models can be applied to condensed matter systems with a degree of confidence
that approaches that of their conformally invariant (non-supersymmetric) counterparts.
In the following we will use a relatively simple gravitational model to illustrate the holographic
approach. The first step is to realize the asymmetric scaling symmetry (\ref{eq:anisotropic}) as
an isometry of a higher dimensional spacetime. This is achieved by introducing an extra radial
dimension and considering the
$d+2$ dimensional Lifshitz geometry\cite{Koroteev:2007yp,Kachru:2008yh}
\beq
ds^2=L^2(-r^{2z}dt^2+r^2d\textbf{x}^2+\frac{dr^2}{r^2}),
\label{eq:lifshitzmetric}
\eeq
where $L$ is a characteristic length scale that we set to $L=1$ from now on.

The Lifshitz metric is invariant under the transformation
\beq
t\rightarrow \lambda^z\,t , \qquad \vec{x}\rightarrow \lambda \vec{x}, \qquad r\rightarrow
\frac{r}{\lambda} ,
\label{eq:lifshitzscaling}
\eeq
which includes (\ref{eq:anisotropic}) acting on the $d+1$ coordinates $t,\vec{x}$ of the dual field
theory. The scaling acts inversely on the radial coordinate compared with the transverse spatial
coordinates. Under the duality the asymptotic large $r$ region corresponds to short distance UV
physics in the field theory while low-energy IR physics is encoded at small $r$.
So far, this is in direct analogy to the usual AdS/CFT correspondence where the conformal
symmetry of a relativistic field theory is realized as the isometry of $AdS_{d+2}$. Indeed, the
Lifshitz metric \eqref{eq:lifshitzmetric} reduces to that of AdS spacetime in Poincar\'e coordinates
when $z=1$. The Lifshitz metric for $z\neq1$ is singular at $r=0$. All 
curvature invariants remain finite but tidal forces diverge as $r\rightarrow 0$.
This is a puzzling feature but it does not pose any immediate problem for condensed matter 
physics applications as real world systems always have a non-vanishing temperature and 
then the singularity is cloaked by the event horizon of a black brane. 

To develop the analogy further, one looks for a gravitational model which has the Lifshitz metric for
generic $z\geq 1$ as a solution of its field equations. There are a few different options available but
the following Einstein-Dilaton-Maxwell (EDM) theory turns out to be a convenient choice,\cite{Taylor:2008tg}
\beq
S_\textrm{EDM} = \int\mathrm{d}^4x\sqrt{-g}\left(
R-2\Lambda-\frac{1}{2}\partial_\mu\phi\partial^\mu\phi
-\frac{1}{4}\sum_{i=1}^2 e^{\lambda_i\phi}F^{(i)}_{\mu\nu}F^{(i)\mu\nu}\right).
\label{eq:action1}
\eeq
Here we have set $d=2$ to obtain a dual description of a 2+1 dimensional field theory. The
formulas generalize to other values of $d$ in a straightforward way.

The Lifshitz metric (\ref{eq:lifshitzmetric}), with $z$ related to the
cosmological constant through
$\Lambda=-(z+2)(z+1)/2$,
is a solution of the model when the dilaton field and one of the
$U(1)$ gauge fields have the following background values,
\beq
e^\phi = \left(\frac{r}{r_0}\right)^{2\sqrt{z-1}} , \quad
F^{(1)}_{rt}=\sqrt{2(z-1)(z+2)}r_0^{z-1}\left(\frac{r}{r_0}\right)^{z+1}. \label{eq:lifshitzbackground}
\eeq
Here $r_0$ is an arbitrary reference value of the radial variable. The role of the auxiliary gauge
field, $F^{(1)}_{\mu\nu}$, is to modify the asymptotic behavior of the metric from that of AdS spacetime
to Lifshitz via the gravitational back-reaction to its background value. In the limit $z\rightarrow1$ the
dilaton becomes independent of $r$ and the auxiliary gauge field vanishes.

The gauge coupling parameter
is required to be $\lambda_1=-2/\sqrt{z-1}$, which means that $F^{(1)}_{\mu\nu}$ is strongly coupled
in the asymptotic $r\rightarrow\infty$ region. This strong coupling is not of immediate concern as long
as $F^{(1)}_{\mu\nu}$ only couples to the gravitational sector and not to matter fields and we are only
interested in classical solutions of the model. Furthermore, the logarithmic form of the dilaton field in
(\ref{eq:lifshitzbackground}) breaks the scale symmetry (\ref{eq:lifshitzscaling}) unless the scale
transformation is generalized to include a shift in the dilaton. We minimize the effect of this breaking
of scale invariance by only considering observables that do not directly couple to the dilaton.

In order to model finite charge density in the dual field theory the EDM theory contains a second $U(1)$
gauge field that couples to charged matter in the bulk gravitational spacetime. Under the duality,
scalar and spinor fields, that are charged under the second $U(1)$ in the EDM theory, correspond to
charged operators in the dual field theory.

\section{Thermodynamics}

Finite temperature is introduced by constructing static black hole solutions, or more precisely
black branes with a planar event horizon, whose Hawking temperature corresponds to the
temperature of the dual field theoretic system. For $\lambda_2=\sqrt{z-1}$, the field equations
obtained from \eqref{eq:action1} have the following one-parameter family of electrically charged
black brane solutions,\cite{Tarrio:2011de}
\beq
ds^2=-r^{2z}f(r)dt^2+r^2d\textbf{x}^2+\frac{dr^2}{r^2 f(r)},
\label{eq:lifshitzbrane}
\eeq
with
\begin{eqnarray}
f(r)&=&1-\left(1+\frac{\rho_2^2}{4z} \right) \left(\frac{r_0}{r}\right)^{z+2}
+\frac{\rho_2^2}{4z} \left(\frac{r_0}{r}\right)^{2z+2} , \label{eq:branemetric} \\
F^{(2)}_{rt}&=&\rho_2\, r_0^{z-1} \left(\frac{r_0}{r}\right)^{z+1} , \label{eq:braneF2}
\end{eqnarray}
and the dilaton and $F^{(1)}_{rt}$ having the same background values as in
(\ref{eq:lifshitzbackground}). As expected, the metric and the physical gauge field reduce to
those of a standard AdS-Reissner-Nordstr\"om black brane when $z=1$.

The scalar potential of the physical gauge field is obtained by integrating the
field strength in \eqref{eq:braneF2} with respect to $r$,
\beq
A^{(2)}_t=\mu -\frac{\rho_2\, r_0^z}{z}\left(\frac{r_0}{r}\right)^z.
\label{eq:branepotential}
\eeq
According to the AdS/CFT prescription for the generating functional in the dual field theory,
the boundary value of $A^{(2)}_\mu$ acts as the source for the corresponding $U(1)$ current.
The integration constant $\mu$ therefore plays the role of a chemical potential in the
field theory. In order for the gauge connection to be regular at the horizon the scalar potential
must go to zero at $r=r_0$, and as a result the chemical potential and the charge density of a
Lifshitz black brane are related, $\mu=\rho_2r_0^z/z$.

The Hawking temperature is determined in the usual manner by considering the
Euclidean version of the black brane metric and requiring smoothness at the horizon,
\beq
 T=\frac{r_0^{z+1}}{4\pi} f'(r_0) = \frac{r_0^z}{4\pi}\left(z+2-\frac{\rho_2^2}{4}\right).
 \label{eq:temperature}
\eeq
The metric \eqref{eq:branemetric} is invariant under a Lifshitz rescaling \eqref{eq:lifshitzscaling}
combined with $r_0\rightarrow \lambda^{-1}r_0$, but the value of the Hawking temperature
scales under this transformation. Due to the underlying scale invariance of the
system, the Hawking temperature by itself does not have physical meaning but only
dimensionless combinations such as $T/\mu$.

The free energy of the dual field theory is obtained from the gravitational theory
by computing the on-shell Euclidean action, including the Gibbons-Hawking boundary
term\cite{Gibbons:1976ue} and boundary counterterms for holographic renormalization.\cite{Henningson:1998gx,Balasubramanian:1999re,Tarrio:2012xx}
For the charged Lifshitz black brane \eqref{eq:branemetric} we find a scaling form,
\beq
 F=V T^{(z+2)/z} g(T/\mu),
 \label{eq:fscaling}
\eeq
that generalizes the $F\propto V T^3$ scaling found for $z=1$ branes in 2+1 dimensions, 
where $V$ is the volume in the field theory, or rather an area since we are considering a theory
in 2+1 dimensions.

If the gravitational theory has more than one Euclidean solution then the one with lowest
free energy dominates. We will see examples below where the system makes a phase transition
from one type of solution to another when $T/\mu$ crosses a critical value.

\section{Geometric probes}\label{sec:geometricprobes}
Different types of minimal surfaces ending on the boundary of the spacetime provide a 
natural set of geometric observables that correspond to entanglement entropy and Wilson loops
in the dual field theory.
Entanglement entropy in a $d+1$ dimensional field theory is obtained by dividing the $d$ dimensional
space, on which the theory is defined, into several parts and then taking a trace of the density matrix 
over the quantum mechanical Hilbert space of the local degrees of freedom in some of the parts.
This leads to a reduced density matrix for the remaining system that generically has
non-vanishing entropy $S=-\textrm{Tr}\rho\log\rho$, which measures the degree of entanglement
between different spatial parts of the system \cite{Srednicki:1993im,Calabrese:2009qy}. The
entanglement entropy depends on the state of the system and can be computed both in pure 
and mixed states. 

In the following we consider a special case where the $d$ dimensional spatial boundary in the
dual gravity theory is divided into two parts $A$ and $B$. The holographic entanglement 
entropy\cite{Ryu:2006bv} between $A$ and $B$ is then given by the area of the $d$ dimensional 
minimal surface that ends on the $d-1$ dimensional perimeter between the two regions,
\beq
S_{ent}=\frac{1}{4 G_N}\int d^{d}\sigma\sqrt{\textrm{det}_{ab}
\Big(g_{\mu\nu}\frac{\partial x^{\mu}}{\partial\sigma^a}\frac{\partial x^{\nu}}{\partial\sigma^b}\Big)}.
\label{afunctional}
\eeq
From the geometric point of view the holographic entanglement entropy can be regarded as a 
generalization of the Bekenstein-Hawking entropy \cite{Casini:2011kv,Lewkowycz:2013nqa}.
As a concrete example, we consider the Lifshitz spacetime with $d=2$ at zero temperature and 
take $A$ to be a disk of radius $a$ in the transverse $\vec x$ plane. The minimal surface 
inherits the rotational symmetry of the disk and can be parametrized by a single function $u(\rho)$, 
where $\rho$ is a polar coordinate on the spatial
boundary, $u\equiv 1/r$, and $u(a)=0$.  The area functional becomes
\beq
S_{ent}=\frac{\pi}{2G_N}\int_0^a d\rho\,\frac{\rho}{u^2} \sqrt{1+\left(\frac{du}{d\rho}\right)^2},
\eeq
which coincides with the corresponding area functional in AdS$_4$ 
since the spatial metric components of Lifshitz spacetime are the same as those of AdS 
in Poincare coordinates.
The resulting Euler-Lagrange equation is solved by
$u(\rho)=\sqrt{a^2-\rho^2}$ and one finds that the dual field theory with
Lifshitz scale-invariance has the same entanglement entropy for a disk of radius $a$ as a 
$2+1$ dimensional CFT,\cite{Ryu:2006bv}
\beq
S_{ent}=\frac{\pi}{2G_N}\left(\frac{a}{\epsilon}-1 \right).
\eeq
Here $\epsilon$ is a short distance 
cutoff in the dual field theory, related to a long distance cutoff in the radial direction in the 
gravitational theory. The first term is divergent in the limit as the cutoff is taken to zero and 
corresponds to the well known `area law' of entanglement entropy \cite{Srednicki:1993im}. 
It is accompanied by a finite term, which is independent of $a$ and thus
manifestly scale invariant.

At finite temperature the Lifshitz spacetime is replaced by the planar black brane \eqref{eq:lifshitzbrane}. 
Let us again consider the holographic entanglement entropy of a disk of radius $a$ in 
the transverse plane. The area functional depends on the black brane metric but at low temperature, 
{\it i.e.} when $a r_0\ll 1$, the minimal surface remains well separated from the event horizon of the brane 
and the holographic entanglement entropy depends only weakly on the temperature. At high temperature, 
when $a r_0\gg 1$, it instead becomes advantageous for the minimal surface to be as close as possible to 
the brane horizon. It approaches a limiting form that consists of a cylinder,
$\vert\vec x\vert=a$, extending from the spatial boundary to the brane horizon, 
and a disk, $\vert\vec x\vert<a$, at $r=r_0$. The holographic entanglement entropy becomes
\beq
S_{ent}= \frac{\pi a}{2 G_N\epsilon}+\frac{A}{4G_N}+\ldots ,\label{eq:thermalentanglement}
\eeq
where $A=\pi a^2r_0^2$ is the proper area covered by the disk at the horizon and we have  
neglected finite terms that are linear in $a$.
Thus the entanglement entropy of a large region, or equivalently at high temperature, contains the same
cutoff dependent term as we saw at zero temperature and an additional term that equals the thermal 
entropy contained in that region \cite{Ryu:2006bv}. This is physically reasonable from the dual field 
theory perspective. The divergent short distance entanglement across the perimeter of the region is 
not affected by the finite temperature but the remaining finite part of the entropy is no longer scale 
invariant and is given by the thermal entropy.

Next we consider Wilson loops in a gauge theory with a gravity dual. These are gauge invariant observables 
located on a loop in the spacetime of the field theory. If the loop has a rectangular shape extending in the time 
direction, the Wilson loop gives the potential energy $V(l)$ between charges in the fundamental and the 
anti-fundmental representation of the gauge group\footnote{One can also consider Wilson loops in other 
representations of the gauge group.}
through the identification
\beq
\langle W\rangle\approx e^{-iT V(l)},
\eeq
where $l$ is the spatial length of the rectangle and $T$ is its temporal length.

In holography, the expectation value of a Wilson loop is given by the worldvolume of a classical string ending 
on the corresponding loop at the spacetime boundary \cite{Maldacena:1998im,Alday:2007he}.
Strictly speaking, to justify the Wilson loop formula, one should consider duality where the gravitational theory 
is a full fledged string theory. Here we will continue on a more phenomenological path and assume the 
Wilson loop to be given by the on-shell value of the Nambu-Goto action,
\beq
\langle W\rangle\approx e^{iS_{NG}},\quad S_{NG}=-\frac{1}{\alpha'}\int 
d^2\sigma \sqrt{-\textrm{det}_{ab}
\Big(g_{\mu\nu}\frac{\partial X^{\mu}}{\partial\sigma^a}
\frac{\partial X^{\nu}}{\partial\sigma^b}\Big)}.
\label{ngaction}
\eeq
As an example we consider a rectangular region with large temporal length $T$ in the 
Lifshitz spacetime. Parametrizing the worldsheet by
$r=r(x)$ gives
\beq
S_{NG}=-\frac{1}{\alpha'}T\int dx \sqrt{r^{2z+2}+r^{2z-2}(\partial_x r)^2)}.
\eeq
From the conserved "Hamiltonian" $H=\frac{\delta L}{\delta(\partial_xr)}\partial_xr-L$, one 
obtains a first order differential equation for $r(x)$ that can be integrated to give
\beq
l=´2\int_{r*}^{\infty}\frac{dr}{r^2\sqrt{\Big(\frac{r}{r_*}\Big)^{2z+2}-1}}=\frac{2}{r_*}
\int_1^{\infty}\frac{ds}{s^2\sqrt{s^{2z+2}-1}},
\eeq
where $r_*$ is the midpoint of the hanging string.
Similarly the action of the string evaluated on the solution is given by
\beq
S_{NG}=-\frac{2T}{\alpha'}r_*^z\int_1^{1/(\epsilon r_*)}\frac{ds s^{2z}}{\sqrt{s^{2z+2}-1}}.
\eeq
The resulting energy of a charge-anti charge configuration in the dual field theory is
\beq
V(l)=\frac{a}{\epsilon^z}-\frac{1}{\alpha'}\frac{c}{l^z}, 
\label{eq:vacuumloop}
\eeq
where $a$ and $c$ are ($z$ dependent) constants\cite{Danielsson:2009gi}. The first 
term is a divergent self-energy while the second term is the potential
energy between the charges as a function of their separation, with a functional form consistent 
with scale invariance.

One can also compute the Wilson loop at finite temperature by considering a black brane 
background. When $l$ is small, the string worldvolume is located at 
large $r$ giving rise to a potential that is close to the vacuum result (\ref{eq:vacuumloop}). 
When $l$ is increased above a critical value $l_c\propto 1/\sqrt{T}$, the lowest energy 
string solution consists of a pair of vertical strings penetrating the black hole horizon.
This results in screening at distances above a critical value\cite{Danielsson:2009gi}, 
analogous to the $z=1$ case.\cite{Brandhuber:1998bs,Rey:1998bq}

\section{Hyperscaling violation}

An interesting generalization of the Lifshitz geometry is provided by the 
so-called hyperscaling violating metrics,\cite{Gouteraux:2011ce,Huijse:2011ef}
\beq
ds^2=r^{-2\theta/d}\Big(-r^{2z}dt^2+r^2d\textbf{x}^2+\frac{dr^2}{r^2}\Big),\label{eq:hyperscaling}
\eeq
in $d+2$ dimensions, which are covariant $ds^2\rightarrow \lambda^{2\theta/d}ds^2$ under the scaling
$r\rightarrow \lambda^{-1}r$, $t\rightarrow \lambda^z t$ and $\textbf{x}\rightarrow \lambda\textbf{x}$,
when $\theta$ is non-vanishing. 
Such metrics can appear, for instance, in Einstein-Maxwell-Dilaton theories 
when a potential for the dilaton field is included \cite{Huijse:2011ef} and also in the context of 
holographic superfluids in the zero temperature limit when the scalar and gauge fields in the 
bulk gravitational theory have non-minimal couplings\cite{Gouteraux:2012yr}. We will have more
to say about holographic superfluids in Section~\ref{holosuperfl} below.

The hypescaling violating metric (\ref{eq:hyperscaling}) has a naked singularity at $r=0$ but at 
finite temperature it is cloaked by the event horizon of a black brane. 
We consider a specific generalization of the EMD theory (\ref{eq:action1}) 
with a potential $U(\phi)=U_0e^{\gamma\phi}$ for the dilaton.\cite{Alishahiha:2012qu}
This theory has neutral black brane solutions given by
\beq
ds^2=r^{-2\theta/d}\Big(-r^{2z}f(r)dt^2+r^2d\textbf{x}^2+\frac{dr^2}{f(r)r^2}\Big)\,,\quad 
f(r)=1-\Big(\frac{r_0}{r}\Big)^{z+d-\theta}.\label{eq:hyperscaling2}
\eeq
The Hawking temperature is obtained as in (\ref{eq:temperature}) and scales
as $T\propto r_0^{z}$. The entropy of the black brane is proportional to the horizon area
$S\propto V_d r_0^{d-\theta}$, where $V_d$ is the spatial volume of the field theory, or 
when expressed in terms of the temperature, 
\beq
S\propto V_d T^{\frac{d-\theta}{z}}.
\eeq
This differs from the behavior of a scale invariant system, where the entropy scales as 
$S\propto V_d T^{d/z}$. In particular, for $\theta=d{-}1$, the entropy scales with temperature 
as if the relevant low energy degrees of freedom lived in one dimension. 
This is precisely what one expects to find for the entropy in a system 
with a Fermi surface, where the low energy excitations are located near a $d{-}1$ 
dimensional surface in momentum space and only disperse along the direction orthogonal 
to that surface.\cite{Huijse:2011ef} 

Another hint of a Fermi surface at $\theta=d-1$ comes from the holographic entanglement entropy.
Consider a disk shaped region in $d=2$, as in section \ref{sec:geometricprobes}, but 
this time at the spatial boundary of a $4$-dimensional, zero temperature, hyperscaling 
violating spacetime \eqref{eq:hyperscaling}. The area functional \eqref{afunctional} is given by
\beq
S_{ent}=\frac{\pi}{2G_N}\int_0^a d\rho\,\rho\, u^{\theta-2} \sqrt{1+\left(\frac{du}{d\rho}\right)^2},
\label{thetafunctional}
\eeq
with $\rho$ and $u(\rho)$ defined as in section \ref{sec:geometricprobes}. For the case of
interest, $\theta=d-1=1$, the holographic entanglement entropy exhibits an 
$a \log{(a/\epsilon)}$ form, where $a$ is the radius of the disk in the transverse $\vec x$ plane
and $\epsilon$ is a short distance cutoff in the dual field theory. The $a\log{a}$ behavior is 
characteristic of a system with a Fermi surface.\cite{Ogawa:2011bz,Huijse:2011ef,Wolf,Swingle:2009bf}.

At $\theta=1$ the Euler-Lagrange equation takes the form
\beq
\frac{\frac{d^2u}{d\rho^2}}{1+(\frac{du}{d\rho})^2}+\frac{1}{\rho}\frac{du}{d\rho}+\frac{1}{u}=0,
\label{thetaoneeq}
\eeq
and one looks for a solution $u(\rho)$, such that $u(a)=0$ and $u(\rho)>0$ for 
$0\leq \rho<a$. Unlike the $\theta=0$ case, we are not able to find a closed form solution 
to the $\theta=1$ equation, but it is straightforward to 
obtain a numerical solution starting from initial data $u(0)=1$, $u'(0)=0$. 
As expected, the numerical solution is concave and hits $u=0$ at some finite
$\rho=\rho_0$. Noting that equation \eqref{thetaoneeq} is invariant under 
$\rho\rightarrow \alpha\,\rho$, $u\rightarrow \alpha\,u$, a new solution that satisfies
$u(a)=0$ is obtained by choosing $\alpha =a/\rho_0$. 

The leading behavior of the holographic entanglement entropy, including the UV divergent part,
is determined by the small $u$ asymptotics near the spatial boundary and these can be 
obtained analytically as follows, without having to rely on a numerical solution.\cite{friedan}
Introducing a new dimensionless variable $s$ through
\beq
\frac{1}{s}= \frac{u}{a}\sqrt{1+\left(\frac{du}{d\rho}\right)^2},
\label{sdefinition}
\eeq
and viewing $\rho$ and $u$ as functions of $s$, allows us to rewrite the Euler-Lagrange 
equation \eqref{thetaoneeq} as
\beq
\rho\frac{d\rho}{ds}=s\left(\frac{du}{ds}\right)^2.
\label{thetaoneeq2}
\eeq
The previous two equations can be re-expressed as
\beq
\hat \rho =-\frac{\sqrt{1-s^2\hat u^2}}{\hat u} \frac{d\hat u}{ds} \,,\qquad
\frac{d\hat\rho}{ds}=-\frac{s\hat u}{\sqrt{1-s^2\hat u^2}} \frac{d\hat u}{ds} \,,
\label{rhoeqs}
\eeq
where $\hat \rho=\rho/a$ and $\hat u=u/a$. At the spatial boundary $\hat \rho\rightarrow 1$
and $\hat u=\alpha e^{-s}+\ldots$, with $\alpha>0$ a constant. The limit is approached as 
$s\rightarrow \infty$ and near the boundary the pair of equations \eqref{rhoeqs} can be 
solved order by order in $e^{-s}$. The solution takes the form
\begin{eqnarray}
\hat u(s) &=& \alpha e^{-s}+\sum_{k=1}^\infty \alpha^{2k+1}e^{-(2k+1)s} u_k(s)\,, \nonumber\\
\hat\rho(s) &=& 1-\sum_{k=1}^\infty \alpha^{2k}e^{-2ks} \rho_k(s) \,. 
\end{eqnarray}
The $u_k(s)$ and $\rho_k(s)$ are universal polynomials in $s$, independent of $a$ and $\alpha$,
that are determined recursively from \eqref{rhoeqs},
\beq
u_1(s)=\frac{1}{4}s^2-\frac{1}{8}, \quad \rho_1(s)=\frac{1}{2}s+\frac{1}{4},\quad\ldots .
\eeq
The area functional \eqref{thetafunctional} can be expressed as
\beq
S_{ent}=\frac{\pi a}{2G_N}\int_{s_\textrm{min}}^{s_\textrm{max}} ds \left(\frac{1}{\hat u}\frac{d\hat u}{ds}\right)^2,
\eeq
where $s_\textrm{min}$ and $s_\textrm{max}$ are the values of $s$ at $\rho=0$ and $u=\epsilon$ respectively,
and we have introduced a short distance cutoff in the field theory, as in section \ref{sec:geometricprobes}. 
For $\epsilon\ll a$ the upper limit of the $s$ integration is at $s_\textrm{max}=\log{(a/\epsilon)}+\log\alpha+\ldots$
and at large $s$ the integrand reduces to 1 at leading order in $e^{-s}$. The holographic entanglement 
entropy is thus given by
\beq
S_{ent}=\frac{\pi}{2G_N}\left( a \log{a}-a\log{\epsilon}+\ldots\right) ,
\label{finalentropy}
\eeq
as promised.
This result is universal in the sense that the free parameter $\alpha$ in the solution for $\hat u$ and $\hat\rho$
does not enter in the leading $a \log{a}$ behavior of the finite part of the entropy but only in subleading
terms, indicated by $\ldots$ in \eqref{finalentropy}.

\section{Coupling to scalar matter}

\subsection{Probe fields}\label{sec:scalarcorrelator}

Correlation functions of local operators play a central role in gauge theory - gravity duality.
Gauge invariant local operators involving a single trace over the gauge theory color indices 
are dual to local fields in the bulk gravitational theory.\footnote{In the string theory context this 
identification assumes that the string length $\sqrt{\alpha'}$ is small compared to the characteristic 
length scale of the spacetime. In this case, states that correspond to string excitations can be 
approximated by local fields.} 

As an example, consider a free scalar field governed by the action
\beq
S=-\frac{1}{2}\int d^{4}x\sqrt{-g}\Big((\partial\varphi)^2+m^2\phi^2\Big).
\label{eq:scalaraction}
\eeq
Two point correlation functions of the corresponding dual operator are obtained by solving
the classical equation of motion of the bulk field, $(\Box-m^2)\phi=0$. 
The solutions have the asymptotic form
\beq
\phi(r,\omega,\textbf{k})=A(\omega,\textbf{k})r^{-\Delta_-}+B(\omega,\textbf{k})r^{-\Delta_+}+...,
\label{eq:asymptotics}
\eeq
where
\beq
\Delta_{\pm}=\frac{2+z}{2}\pm \sqrt{\frac{(2+z)^2}{4}+m^2},
\label{eq:scalingdimension}
\eeq
and we have performed a Fourier transform 
$\varphi(x,t)=\int \frac{d^{2}kd\omega}{(2\pi)^{3}}e^{-i\omega t+i \textbf{k}\cdot\textbf{x}}\phi(r,\omega,\textbf{k})$. 
The leading term $A(\omega,\textbf{k})$ is identified as a source for the operator dual to the field $\phi$, 
while $B(\omega,\textbf{k})$ is identified as the expectation value of the dual operator 
$\langle \mathcal{O}\rangle$. 
In linear response theory one identifies the retarded correlation function as the ratio of the expectation 
value and the source, \cite{Son:2002sd,Herzog:2002pc,Iqbal:2008by}
\beq
G_R(\omega,\textbf{k})\propto \frac{\langle\mathcal{O}(\omega,\textbf{k})\rangle}{A(\omega,\textbf{k})}\propto \frac{B(\omega,\textbf{k})}{A(\omega,\textbf{k})}.\label{eq:retarded}
\eeq
The retarded correlator describes the response of the system to turning on a source. 
It is a causal quantity and vanishes in the past of the earlier operator inside the expectation value.
One is thus lead to impose ingoing boundary conditions with vanishing boundary data on 
any past horizons that are present in the spacetime. The boundary conditions fix $B(\omega,\textbf{k})$
in terms of $A(\omega,\textbf{k})$ and thereby determine the correlation function \eqref{eq:retarded}
up to an overall normalization.
It is straightforward to generalize the above procedure to interacting scalar fields and fields other 
than scalars. The bulk field equations can be also solved in position space, leading directly to 
correlation functions in position space in the dual field theory. 

In the limit of large scalar field mass, the path integral for the two point correlation function
can be evaluated in a saddle point approximation,\cite{Hartle:1976tp,Balasubramanian:1999zv}
\beq
G(x_2,x_1)=\langle\mathcal{O}(x_2)\mathcal{O}(x_1)\rangle\propto e^{iS_{cl}},
\eeq
where $S_{cl}$ is the classical action of a relativistic particle of mass $m$ along a geodesic 
connecting the two boundary points $x_1$ and $x_2$,
\beq
S_{cl}=m\int d\lambda \sqrt{-g_{\mu\nu}\frac{dx^\mu}{d\lambda}\frac{dx^\nu}{d\lambda}} .
\label{eq:geodesicaction}
\eeq
The geodesic approximation provides a simple way to calculate position space correlators. 
To illustrate its use, we consider the equal time correlation function of scalar operators in
the field theory dual to the Lifshitz spacetime \eqref{eq:lifshitzmetric}, where spacelike 
geodesics satisfy
\beq
\frac{dt}{d\lambda}=\frac{E}{r^{2z}} ,\qquad
\frac{dx}{d\lambda}=\frac{P}{r^{2}} ,\qquad
\left(\frac{dr}{d\lambda}\right)^2=r^2-P^2+E^2r^{2-2z} .
\eeq
Here $E$ and $P$ are constants (for the equal time correlator we set $E=0$) and we have 
used translation and rotation invariance in the $\vec x$-plane to align the geodesic along 
the $x$-axis.
The geodesic equations are easily integrated and the resulting action is
\beq
S=im(\lambda_2-\lambda_1)=2im\log\left(|x_2-x_1|/\epsilon\right) ,
\eeq
where $\lambda_1$ and $\lambda_2$ are the values of the affine parameter at the  
endpoints of the geodesic and we have cut off the integral at $r=1/\epsilon$. 
The action takes an imaginary value since the equal time geodesic is spacelike. 
This leads to the correlation function
\beq
\langle\mathcal{O}(x_1)\mathcal{O}(x_2)\rangle\propto\frac{1}{|x_2-x_1|^{2\Delta}},
\label{eq:equaltimecorrelator}
\eeq
with $\Delta=m$. The algebraic decay of the correlation is a sign of scale invariance 
at a quantum critical point. It goes beyond the geodesic approximation and 
\eqref{eq:equaltimecorrelator} holds for a scalar field of any mass, with $\Delta$ given 
by \eqref{eq:scalingdimension}.\cite{Kachru:2008yh}

Next we consider the equal time correlation function in a non-extremal Lifshitz black 
brane background \eqref{eq:lifshitzbrane}. An equal time geodesic connecting two
well separated boundary points $x_1$ and $x_2$, such that $\vert x_2-x_1\vert r_0\gg1$,
consists two `vertical' segments $r_0<r<1/\epsilon$ at $x=x_1$ and $x=x_2$, respectively, 
connected by a `horizontal' segment along the horizon at $r=r_0$. The classical 
action \eqref{eq:geodesicaction} of such a geodesic is given by
\beq
S_{cl}=2im\log{\left(\frac{1}{\epsilon}\right)}+i m r_0 \vert x_2-x_1\vert+\ldots ,
\eeq
where $\ldots$ are finite terms that do not depend on $\vert x_2-x_1\vert$.
This leads to an exponentially decaying correlation ,
\beq
G(x_2,x_1)\propto e^{-m r_0\vert x_2-x_1\vert },
\eeq
characteristic of a thermal system with a correlation length $\xi =1/(m r_0)$. 
This result holds quite generally as long
as the metric function $f(r)$ in \eqref{eq:branemetric} has a simple zero at $r=r_0$.
The geodesic approximation breaks down for small scalar field masses but numerical
calculations confirm that equal time correlation functions have a thermal character
in this case as well.\cite{Keranen:2012mx}

Two-point correlation functions involving operators at timelike separated boundary 
points are more difficult to calculate as they arise from quantum tunneling. 
In particular, there are no real valued geodesics that connect the boundary points 
in this case. The geodesic approximation can still be applied, using 
Euclidean time methods and analytically continuing the answers to 
real time at the end, provided the 
Euclidean result is known in closed form.\cite{Balasubramanian:2012tu}
More generally, timelike correlators can be computed directly in real time by solving 
the scalar field equations following from  
(\ref{eq:scalaraction}) for a general diagonal and
time independent metric,
\beq
\frac{1}{\sqrt{-g}}\partial_r(\sqrt{-g}g^{rr}\partial_r\phi)-(\textbf{k}^2g^{xx}+\omega^2 g^{tt}+m^2)\phi=0.\label{eq:scalareq}
\eeq
In the simple example of a BTZ black brane in $2+1$ dimensions, with
$f(r)=1-\frac{r_0^2}{r^2}$,
the scalar field equation (\ref{eq:scalareq}) has a closed form solution in terms 
of hypergeometric functions, from which it is straightforward to calculate the retarded 
correlation function using (\ref{eq:retarded}). The spectral function,
which is the imaginary part of the retarded correlator, is given by\cite{Son:2002sd}
\beq
\textrm{Im}(G_R)\propto \sinh\Big(\frac{\omega}{2T}\Big)|\Gamma\Big(\frac{\Delta}{2}
-\frac{i}{4\pi T}(\omega-k)\Big)\Gamma\Big(\frac{\Delta}{2}-\frac{i}{4\pi T}(\omega+k)\Big)|^2.
\label{eq:btz}
\eeq
The spectral function only has poles in the lower half of the complex $\omega$ plane,
\beq
\omega=\pm k-4\pi i T(\frac{\Delta}{2}+n),  \qquad  n\in\{0,1,2,\ldots\},
\eeq
so the correlation function decays exponentially in time. Small perturbations away from 
thermal equilibrium decay on a time scale $t=1/(2\pi T\Delta)$, determined by the pole 
at which $\omega$ has the smallest imaginary part. 
The poles of the correlation function correspond to solutions, called quasinormal modes,
for which $A(\omega,k)$ in \eqref{eq:retarded} vanishes.

\subsection{Holographic superfluids}\label{holosuperfl}

As opposed to the neutral BTZ background considered above, quasinormal modes of a scalar 
field can migrate to the upper half complex $\omega$ plane in charged black brane 
backgrounds\cite{Gubser:2008px}. This occurs in many gravity models with a negative 
cosmological constant and indicates an instability of the black brane towards a configuration 
with non-vanishing background scalar field. In other words, black branes can grow scalar hair.  
To see how this may happen, consider a charged scalar field in the charged black brane 
background \eqref{eq:lifshitzbrane},
\beq
(\Box-(m^2+e^2g^{tt}A_t^2))\phi=0,
\eeq
where $e$ is the charge of $\phi$ and $A_t(r)$ is the 
brane gauge field \eqref{eq:branepotential}.\footnote{The scalar field only couples to the 
physical gauge field $A^{(2)}_\mu$ so we drop the $^{(2)}$ superscript.} 
The combination 
\beq
m_\textrm{eff}^2=m^2+g^{tt}A_t^2,
\eeq
can be interpreted as an effective mass squared for the scalar field and if the original scalar field mass 
squared is not too large, the gauge field term may dominate. Since $g^{tt}$ is negative 
outside the brane horizon, the effective mass is then tachyonic and if it violates the 
Breitenlohner-Freedman bound\cite{Breitenlohner:1982jf} one expects an instability 
towards condensation of the scalar field outside the black brane. Near the spacetime boundary 
$r\rightarrow\infty$ the effective mass is dominated by the bare mass term $m^2$, so any
condensate that forms near the brane horizon will fall off at large $r$.

There is another mechanism for a scalar field instability in a charged black brane 
background.\cite{Hartnoll:2008kx} The near-horizon geometry of an extremal black brane in 
asymptotically Lifshitz spacetime is given by AdS$_2\times R^2$. At low
temperatures the brane is near-extremal and the near-horizon region approaches 
AdS$_2\times R^2$. The Breitenlohner-Freedman bound is stronger in 
AdS$_2$ than in the asymptotic region,
\beq
m^2_{BF}\Big\vert_\textrm{AdS$_2$}=-\frac{1}{4}\> 
> \> -\frac{(z+d)^2}{4} =m^2_{BF}\Big\vert_{r\rightarrow\infty}\,,
\eeq
so even a neutral scalar field can be sufficiently tachyonic in the 
near-horizon region to cause condensation at low temperature if the mass squared is in the
above range.

To illustrate the scalar condensation, we consider a large $e$ probe limit where the back reaction
of the scalar field and the gauge field on the brane geometry can be neglected\cite{Hartnoll:2008vx}. 
Further assuming that $d=2$, $z=1$, and $m^2=-2$, the problem is reduced to considering
\beq
S=\frac{1}{e^2}\int d^4x\sqrt{-g}\Big(-\frac{1}{4}F_{\mu\nu}^2-|D_{\mu}\phi|^2+2|\phi|^2\Big),
\label{eq:superaction}
\eeq
in the background geometry of an AdS-Schwarzschild black brane. A time independent 
equilibrium configuration with translation and rotation symmetry in the transverse plane
amounts to an ansatz of the form $\phi=\phi(r)$, $A_t=A_t(r)$. The $r$ component of Maxwell's 
equations requires the phase of $\phi$ to be a constant, which may be chosen so that $\phi$ 
is real valued. The field equations obtained from \eqref{eq:superaction} then reduce to 
\begin{align}
\phi''+\Big(\frac{f'}{f}+\frac{2}{r}\Big)\phi'+\frac{A_t^2}{f^2}\phi+\frac{2}{f}\phi=0,
\\
A_t''+\frac{2}{r}A_t'-2\frac{\phi^2}{f}A_t=0,
\end{align}
where $f(r)=-g_{tt}=g^{rr}$ is the function that appears in the background metric.
The boundary conditions for the fields at the horizon follow from regularity: For $A=A_t(r)dt$ to 
have finite norm we must have $A_t=0$ at the horizon, and, requiring $\phi''$ to be finite, the equation 
of motion of $\phi$ implies $f'\phi'+2\phi=0$ at the horizon. The behavior of the bulk fields near the 
AdS boundary provides input data for the dual field theory. The boundary value of $A_t$ gives the 
field theory chemical potential $\mu$ and we require that the scalar operator that is dual to $\phi$ 
has vanishing source. For the chosen value of $m$ these conditions amount to
\beq
\phi=\frac{\phi_+}{r^2}+O(r^{-3}),\quad A_t=\mu+\frac{\rho}{r}+O(r^{-2}).
\label{eq:bulkasymptotics}
\eeq
The field equations can be solved numerically, subject to these boundary conditions. The 
chemical potential introduces a reference scale and when the Hawking temperature of the black
brane is smaller than a critical temperature $T_c\approx .059\mu$ one finds solutions with a 
non-vanishing scalar field profile satifying the above conditions\cite{Hartnoll:2008vx}. 
The critical value corresponds to the temperature at which the first scalar quasinormal 
mode crosses to the upper half of the complex $\omega$ plane.
Below the critical temperature the scalar field condenses and this spontaneously breaks 
the global $U(1)$ symmetry of the dual field theory. By studying fluctuations around the 
condensate background one finds that it is stable\cite{Amado:2009ts}, 
unlike the normal phase with $\phi=0$ and $A_t=\mu-\rho/r$. 

\begin{figure}[t]
\begin{center}
\psfig{file=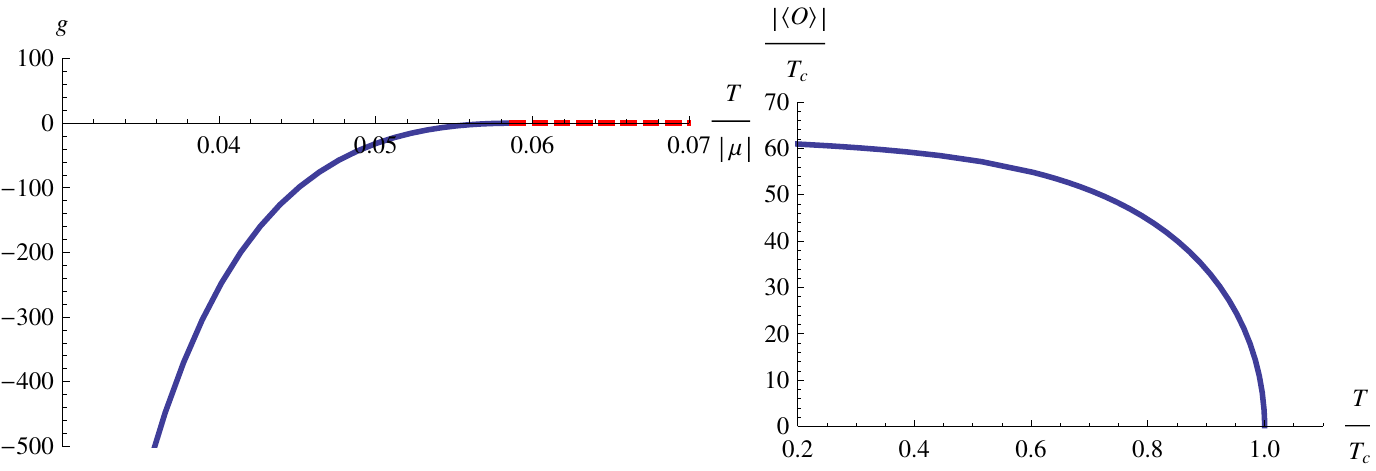,width=5 in}
\end{center}
\caption{ Left: Free energy difference between the condensate and the normal phase. Right: The 
expectation value of the scalar operator dual to $\phi$ in the condensed phase.}
\label{aba:fig2}
\end{figure}

The condensate solution is thermodynamically stable. 
The free energy in the field theory is identified with the Euclidean on shell action of the
bulk system. Since the temperature and the chemical potential are the only scales in the 
problem, the free energy has a scaling form
\beq
\Delta F=F_\textrm{cond}-F_\textrm{normal}=VT^3 g(T/|\mu|),
\eeq
where $V$ is the transverse volume, $F_\textrm{cond}$ is the free energy of the condensate 
solution, and $F_\textrm{normal}$ is the free energy of the normal phase solution. Comparing the 
free energies of the numerical solution for $\phi$ and $A_t$ in the condensed phase and the known
analytic solution for the normal phase leads to the result shown on the left in Figure~\ref{aba:fig2}.
The condensed phase, when it exists, is indeed found to have the lower free energy. 
The phase transition at the critical temperature is seen to be continuous and of second order.

As usual in AdS/CFT the asymptotic behavior of bulk fields near the AdS boundary provides
information about physical quantities in the dual field theory. For instance the coefficients in 
\eqref{eq:bulkasymptotics} determine the one point functions of the $U(1)$ current and the 
scalar operator dual to $\phi$,
\beq
\langle O\rangle\propto \phi_+,\qquad \langle J^0\rangle=\rho.
\eeq
The expectation value of the scalar operator $\mathcal{O}$ is shown on the right in 
Figure~\ref{aba:fig2}. Fitting the numerical data near the critical temperature yields
\beq
\langle O\rangle\propto (T_c-T)^{\beta},\label{eq:meanfield}
\eeq
with $\beta\approx 1/2$, consistent with mean field scaling at a 
second order phase transition.

The above scalar field instability is not restricted to the probe approximation and the main conclusions 
for the model with $d=2$ and $z=1$ remain unchanged when the 
back reaction of the scalar field and the gauge field is taken into account.\cite{Hartnoll:2008kx}
Qualitatively similar results are found for other values of $d$,\cite{Horowitz:2008bn} and more
general $z$.\cite{Brynjolfsson:2009ct}
The nature of the phase transition is sensitive to details of the gravitational model. 
Adding strong enough self interactions for the scalar field or a non-minimal gauge field coupling 
can, for instance, change the scaling exponent $\beta$ in \eqref{eq:meanfield} 
and the order of the phase transition can go from second order to first order\cite{Franco:2009yz,Aprile:2009ai}.

The AdS$_2\times R^2$ near-horizon limit of an extremal charged brane without hair is unphysical
in the sense that it corresponds to finite entropy density at zero temperature in the dual field theory.
The condensate solution behaves rather differently at low temperature because of the strong back 
reaction due to the scalar field on the near-horizon geometry. The end result is model dependent
and here we will restrict out attention to simple models of the form\cite{Gubser:2009cg}
\beq
S=\frac{1}{2\kappa^2}\int d^4x\sqrt{-g}\Big(R-\frac{1}{4e^2}F_{\mu\nu}^2-|D_{\mu}\phi|^2-U(|\phi|^2)\Big),\label{eq:gubser}
\eeq
where the potential $U$ is assumed to be bounded from below.\footnote{The case where $U$ 
includes only a mass term and is not necessarily bounded from below has also been considered 
in detail\cite{Horowitz:2009ij}.} 
The spacetime will approach AdS$_4$ in the UV $r\rightarrow\infty$, and then a charged scalar 
condensate builds up towards IR at small values of $r$. Rather than attempting to construct the
full spacetime at zero temperature, we concentrate on the possible IR limits. As $U$ is bounded
from below, it seems reasonable to assume that $\phi$ will settle to a constant value $\phi_0$ at
small $r$. We also assume that the ground state preserves translation and rotation symmetry
in the transverse $\vec x$ plane. The geometry can then be viewed as a domain wall interpolating 
between the UV AdS$_4$ vacuum and an IR vacuum with $\phi=\phi_0$. The scalar field equation 
can be satisfied by a constant field provided that field value minimizes the effective potential,
\beq
U_{eff}=U(|\phi_0|^2)+g^{tt}(r)A_t(r)^2\phi_0^*\phi_0.\label{eq:effective}
\eeq
This can happen in two ways. Either both terms in (\ref{eq:effective}) are separately extremised, or
the variations of the two terms cancel each other.
In the first case $A_t=0$ and $\phi_0$ extremises the potential $U$ leading to an AdS$_4$ spacetime 
in the IR with a cosmological constant that in general differs from the UV value. In the second case,
where the variations of the two terms cancel, the $r$ dependence of $A_t$ has to be 
$A_t\propto \sqrt{-g_{tt}}=g(r)$. The Maxwell equations then imply that $rg'/g$ must be a constant. 
Denoting this constant by $z$ gives $A_t\propto r^z$. Einstein's equations can then be easily solved 
to find
\beq
ds^2=-\Big(\frac{r}{L_0}\Big)^zdt^2+\frac{L_0^2}{r^2}dr^2+\frac{r^2}{L_0^2}d\textbf{x}^2,\quad \phi=\phi_0,\quad A_t=\sqrt{2-\frac{2}{z}}\Big(\frac{r}{L_0}\Big)^z.
\eeq
For a given potential $U$ and scalar field charge $e$ one can determine $z$ and $\phi_0$ from Einstein's 
equations, although real solutions for $z$ and $\phi_0$ might not always exist.
A more detailed analysis\cite{Gubser:2009cg} indicates that for large enough charge $e$, the IR spacetime 
is AdS$_4$,  while for small $e$, the IR spacetime will be a Lifshitz spacetime.
From the field theory perspective, the different IR limits correspond to whether the operator $J^t$ is 
relevant or irrelevant in the renormalization group sense in the IR. When it is irrelevant, conformal 
symmetry including Lorentz invariance emerges in the IR. On the other hand, when $J^t$ is a relevant 
operator, the dual field theory flows to a scale invariant fixed point with Lifshitz symmetry.

Again one can ask how much of this story depends on the particular action we are considering. 
More general models can certainly lead to other zero temperature geometries. In particular, one can 
obtain hyperscaling violating geometries as zero temperature IR limits if one allows non-minimal 
couplings between the gauge field and the scalar field \cite{Gouteraux:2012yr}.

\section{Fermionic matter}

Fermions are ubiquitous in condensed matter physics and by coupling fermions
to the bulk gravitational theory the AdS/CFT prescription can be extended to 
correlation functions of fermion operators in the dual field theory.\cite{Iqbal:2009fd}
Analogous to the scalar case considered in section~\ref{sec:scalarcorrelator}, the retarded 
two point correlation function of operators dual to a bulk fermion of mass $m$ 
and charge $q$ is obtained by solving the Dirac equation,
\beq
\Big(\Gamma^{\mu}D_{\mu}+m\Big)\Psi =0,
\label{diraceq}
\eeq
with ingoing boundary conditions on $\Psi$ at the black brane 
horizon.\cite{Liu:2009dm,Cubrovic:2009ye} We will not describe the calculation here
but simply note some results of a detailed study at finite charge density and zero temperature 
involving $z=1$, $\theta=0$ extremal branes.\cite{Faulkner:2009wj}

The correlation function develops a pole at $\omega=0$ 
for a certain discrete value of the momentum $k=k_F\sim\mu$.  The dual finite density system
then has gapless states on a shell in momentum space, which is a signature of a Fermi surface.
The physics near the Fermi surface depends on the scaling behavior of solutions to \eqref{diraceq}
near the brane horizon through the parameter $\nu=\sqrt{m^2-q^2+\frac{k_F^2}{\mu^2}}$. 
Suppressing spinor indices, the correlation function has the form 
\beq
G_R(\omega,k) \approx \frac{Z(k)}{\omega-\omega_*(k)+i\Gamma(k)},
\label{fermisurface}
\eeq
for $k\approx k_F$. For $\nu>1/2$ we have $\omega_*\approx v_F(k-k_F)$ and 
$\Gamma\sim (k-k_F)^{2\nu}\ll \omega_*$ 
as $k\rightarrow k_F$, indicating the presence of stable quasi-particles. The value of the Fermi 
velocity $v_F\sim\mu$ is determined by the UV physics at $r\rightarrow\infty$. For general parameter
values the width of the holographic quasi-particles differs from the $\Gamma\sim \omega_*^2$ found
in Landau-Fermi liquid theory. Further departure from Fermi liquid behavior is seen for $\nu<1/2$.
In this case, the frequency and width remain comparable as $k\rightarrow k_F$ and the would be 
quasi-particles are unstable. For the special case $\nu=1/2$, the correlation function
is that of a so called marginal Fermi liquid\cite{Varma:1989}, with the quasi-particle width suppressed 
compared to the frequency but only logarithmically while the quasi-particle 
residue $Z$ vanishes logarithmically as $k\rightarrow k_F$. 

So far we have ignored interactions among the bulk fermions and any back reaction on the metric 
and the gauge field. Going beyond this fermion probe approximation rather quickly leads to 
computational complexity rivaling that of the original strongly coupled many body problem
that the gravity dual is supposed to model. As a final topic, we will briefly describe an interesting
set of geometries that arise when the bulk fermions are treated in a Thomas-Fermi 
approximation as a continuous charged fluid\cite{Hartnoll:2009ns,Hartnoll:2010gu}. 
The Compton wavelength of the bulk fermion is assumed to be small compared to the 
AdS length scale, which means that the scaling dimension of the corresponding fermion operator 
in the dual field theory is large, $\Delta\approx m \gg 1$, which is not the range of parameters of 
direct interest for condensed matter applications, but simplifies the bulk fermion problem dramatically.

We set $d=2$, $z=1$ and assume translation and rotation symmetry in the transverse plane. 
The fermions are described as a charged perfect fluid with energy momentum tensor
$T_{\mu\nu}=(\rho+p)u_\mu u_\nu+p g_{\mu\nu}$ and current density $J_\mu=\sigma u_\mu$, where 
$\rho,p,\sigma$, and $u_\mu$ are the local energy density, pressure, charge density and 
four velocity of the fluid, respectively. Using a free fermion equation of state for the fluid takes into
account the fermionic character of the particles but ignores their mutual interactions. 
For static configurations, the coupled system of Einstein, 
Maxwell, and fluid equations reduce to a set of coupled ordinary differential equations in the 
radial variable $r$, in a holographic analogy to the standard Tolman-Oppenheimer-Volkov
equations for stellar structure\cite{deBoer:2009wk}.

At high temperature, or more precisely high $T/\mu$, the only solution with given mass and charge 
density is an AdS-Reissner-Nordstr\"om charged black brane with vanishing fluid density everywhere. 
As the temperature is lowered, at fixed chemical potential, a second solution becomes possible 
where a `cloud' of fermion fluid is suspended over the horizon of a black brane with sharply defined
inner and outer edges at which the fluid density goes to zero.\cite{Puletti:2010de,Hartnoll:2010ik}
The cloud can be viewed as a fermion analog of the scalar hair on the black brane in a holographic
superfluid. The sharp edges of the cloud are an artifact of treating the fermions as a continuous fluid 
and are smoothed out in more quantum mechanical treatments.\cite{Medvedyeva:2013rpa,Allais:2013lha}
Whenever the solution with a fluid cloud outside the brane is allowed it has lower free energy 
than the AdS-Reissner-Nordstr\"om solution at the same temperature. In the fluid approximation, 
the system undergoes a third order continuous phase transition from the black brane to the fermion 
cloud phase\cite{Puletti:2010de,Hartnoll:2010ik}, but the phase transition is first order in a model  
based on a WKB approximation to the Dirac equation for the bulk fermions\cite{Medvedyeva:2013rpa}. 

In the zero temperature limit the fermion cloud expands in the radial direction, the black brane
horizon recedes, and the solution approaches that of an `electron star'\cite{Hartnoll:2010gu}, which
interpolates between AdS$_4$ in the UV at $r\rightarrow\infty$ and four dimensional Lifshitz 
spacetime in the IR at $r\rightarrow 0$ with a dynamical critical exponent that depends on 
model parameters. This is analogous to the emergent IR Lifshitz behavior at zero temperature
seen in certain models of holographic superfluids, as described in section~\ref{holosuperfl} above.

\section*{Acknowledgments}

This work was supported in part by the Icelandic Research Fund and
by the University of Iceland Research Fund. We would like to thank Daniel Friedan, 
Valentina Giangreco M. Puletti, Esko Keski-Vakkuri, and Tobias Zingg 
for useful discussions.


\end{document}